%% file: LCC2006.tex
\documentclass{llncs}

\usepackage{url}

\usepackage{calc,fancyheadings,xspace}
\usepackage[modulo]{lineno}  

\usepackage{amssymb} 
\usepackage{amsmath}  
\usepackage{amsbsy} 
\usepackage{amsfonts} 
\usepackage{euscript} 
\usepackage{graphicx} 
\usepackage{stmaryrd} 

\usepackage{pstricks,pst-node,pst-text,pst-3d,pst-tree}

\renewcommand{\marginparwidth}{10000pt}

\newcounter{hours}\newcounter{minutes}
\newcommand{\printtime}{%
 \setcounter{hours}{\time/60}%
  \setcounter{minutes}{\time-\value{hours}*60}%
  \thehours :\theminutes}

\include{macroSI}

\begin{document}
\pagestyle{headings} 

\author{Jean-Yves Marion \and Romain P\'echoux}
\institute{Loria, Calligramme project, B.P. 239, 
54506 Vand\oe  uvre-l\`es-Nancy Cedex, France, 
and 
\'Ecole Nationale Sup\'erieure des Mines de Nancy, 
INPL, France.
\email{Jean-Yves.Marion@loria.fr}
\email{Romain.Pechoux@loria.fr}
}

\title{Quasi-friendly sup-interpretations} 
\titlerunning{Sup-interpretations}
\date{\today}
\maketitle

\begin{abstract} 
In a previous paper~\cite{MP06}, the sup-interpretation method was proposed as a new tool to control memory resources of first order functional programs with pattern matching by static analysis. Basically, a sup-interpretation provides an upper bound on the size of function outputs. In this former work, a criterion, which can be applied to terminating as well as non-terminating programs, was developed in order to bound polynomially the stack frame size. In this paper, we suggest a new criterion which captures more algorithms computing values polynomially bounded in the size of the inputs. Since this work is related to quasi-interpretations, we compare the two notions obtaining two main features. The first one is that, given a program, we have heuristics for finding a sup-interpretation when we consider polynomials of bounded degree. The other one consists in the characterizations of the set of function computable in polynomial time and in polynomial space.
\end{abstract}

\section{Introduction}
This paper is part of general investigation on program complexity
analysis and, particularly, on first order functional programming static analysis. 
It studies the notion  
of sup-interpretation introduced in~\cite{MP06},
a method that provides an upper bound on the size of every stack frame if the program is
non-terminating, and establishes an upper bound on
the size of function outputs if the program is terminating. 
Basically, a sup-interpretation is 
a \emph{partial} assignment of symbols,  
which ranges over positive real numbers and which gives a bound on the size of
the computed values. We use this notion to develop a criterion which ensures that the size of the values computed by a program verifying this criterion is polynomially bounded in the size of the inputs and which allows to bound polynomially the size of the stack frames whenever the program is not terminating.

The practical issue is to provide the amount of space resources that a program needs during its
execution. This is crucial for at least
many critical applications, and is of real interest in computer
security. 
There are several approaches which are trying to solve the same
problem. The first one is by monitoring
computations. However, the monitor may crash unpredictably by memory leak if it is compiled with the program. The second one, complementary to static analysis, is a
testing-based approach. Indeed, such an approach provides lower bounds on the memory needed. The last approach is type
checking which can be done by a bytecode verifier. 
Our approach is rather distinct and consists in an attempt to control resources by providing resource
certificates in such a way that the compiled
code is safe w.r.t. memory overflow. Similar works have studied by Hofmann~\cite{Hofmann99,HofmannEsop00} and Aspinall and Compagnoni~\cite{Asp03}. 


The sup-interpretation can be considered as some program annotation provided by the
programmer. Sup-interpretations strongly inherit from:
\begin{itemize}
\item The notion of
quasi-interpretation developed by Bonfante, Marion and Moyen
in~\cite{BMM01,BMM05b,MM00,Marion03}. Quasi-interpretation, like sup-interpretation,
provides a bound on function outputs by static analysis for first
order functional programs and allows the programmer to find a bound on the size of every stack frame. 
The paper~\cite{BMM05b} is a comprehensive introduction to quasi-interpretations which, combined with recursive path orderings, allow to characterize complexity classes such as the set
of polynomial time functions or yet the set of polynomial space functions. Like quasi-interpretations, sup-interpretations, were developed with the aim to pay more attention to the algorithmic aspects of complexity
than to the functional (or extensional) one and then it is part of study of the
implicit complexity of programs. But the main interest of sup-interpretation is to capture
a larger class of  algorithms. In fact,  programs computing
logarithm or division admits a sup-interpretation but have no
quasi-interpretation. Consequently, we firmly believe that sup-interpretations, like quasi-interpretations, could be applied to other languages such as resource bytecodeverifier by following the lines of  ~\cite{AmadioCDJ04} or 
language with synchronous cooperative threads as in~\cite{Amadio04}. 

\item The dependency pair methods introduced by Arts and Giesl in~\cite{AG00} which was initially introduced for
proving termination of term rewriting systems automatically. In order to obtain a polynomial space bound, a criterion is
developed on sup-interpretations using the underlying notion of dependency pairs by Arts and Giesl~\cite{AG00}.

\item The size-change principle by Jones et al.~\cite{JonesSCP} which is another method developed for proving program termination. Indeed, there is a very strong relation between termination and computational complexity and, in order to prove both complexity bounds and termination, we need to control the arguments occurring in the recursive calls of a program.
\end{itemize}

Section 2 introduces the first order functional language and its
semantics. Section 3 introduces the syntactical notion of fraternity which is of real interest to control the size of values added by the recursive calls. Section 4 defines the main notions of sup-interpretation
and weight used to bound the size of a program outputs. In section 5, we introduce a criterion, called quasi-friendly criterion, which enlarges, in practice, the class of programs captured by a former criterion, called friendly criterion, of~\cite{MP06} (for example, it captures algorithms over trees whereas the friendly criterion fails). This criterion provides a polynomial bound on the size of the values and the stack frame size computed by a quasi-friendly programs (depending on whether the programs terminate or not). Finally, in a last section, we also compare the notion of sup-interpretation to the one of quasi-interpretation. First, we show that quasi-interpretation is a particular sup-interpretation. As a consequence, we obtain heuristics for the synthesis of sup-interpretations, which consists in finding a sup-interpretation for a given program, as far as far, we consider the set of $\textbf{Max-Poly}$ functions defined to be constant functions, projections, $\max$, $+$, $\times$ and closed by composition. Finally, using former results about quasi-interpretations, we give two characterizations of the sets of functions computable in polynomial time and respectively polynomial space.

\section{First order functional programming}
\subsection{Syntax of programs} 
In this paper we consider a generic first order functional programming language. 
The vocabulary $\Voc = \langle \Var, \Cns,\Op,\Fct \rangle$ is composed of
four disjoint domains of symbols which represent respectively the set of variables, the set of constructor symbols, the set of basic operator symbols and the set of function symbols. The arity of a symbol is the number $n$ of its arguments.
A program $\p$ of our language is composed by a sequence of definitions $\many{\defone}{m}$ which are basically function symbols definitions and which are characterized by the following grammar:
$$
\begin{array}{llll}
\texttt{Definitions} \ni \textit{def\ } & ::= &\ \funone(\many{\varone}{n})= \eqone^{\funone} \\
\texttt{Expression} \ni \eqone & ::= &\ \varone  \ |\ \conone(\many{\eqone}{n})\ | \ 
       \opone(\many{\eqone}{n})\ | \ \funone(\many{\eqone}{n}) \\
       & & \ |\  \mathbf{Case\ } \many{\eqone}{n} \mathbf{\ of\ } 
\overline{\patone_1} \rightarrow \eqone^1 
\ldots 
\overline{\patone_\ell} \rightarrow \eqone^\ell \\ 
\texttt{Patterns} \ni \patone & ::= &\ x\ | \ \conone(\many{\patone}{n})
\end{array}
$$

where $x, x_1,\ldots,x_n$ are variables,
$\conone \in \Cns$ is a constructor symbol,
$\opone \in \Op$ is an operator symbol,
 $\funone \in \Fct$ is a function symbol,
and $\overline{\patone_i}$
 is a sequence of $n$ patterns. Throughout the paper, we extend this notation $\overline{\eqone}$ in a clarity concern for any sequence of expressions $\eqone_{1},\ldots,\eqone_{n}$, for
some $n$ clearly determined by the context.
 
 The $\mathbf{Case}$ operator is a special symbol that allows pattern matching. It is convenient, because it avoids tedious details, to restrict case definitions in such a way 
that an expression involved in a $\mathbf{Case}$ expression does 
not contain nested $\mathbf{Case}$
(In other words, an expression $\eqone^j$ does not contain 
a $\mathbf{Case}$ expression).
This is not a severe
restriction since a program involving nested  $\mathbf{Case}$ can be
transformed in linear time in its size into an equivalent
program without the nested $\mathbf{Case}$ construction.
 
In a definition, a variable of $\eqone^{\funone}$ is either a variable in
the parameter list $\many{\varone}{n}$ of the definition of $\funone$
or a variable which occurs in a pattern of a $\mathbf{Case}$ definition. 
In a $\mathbf{Case}$ expression, patterns are not overlapping. Such a restriction ensures that considered programs are confluent.

\subsection{Semantics}
The computational domain of a program $\p$ is $\ValuesE= \Values \cup \{ \erreur \}$ where $\Values$ represents the constructor algebra $\Const$ and $\erreur$ is a special symbol returned by the program when an error occurs. 
Each operator symbol $\opone$ of arity $n$ is interpreted by a function $\sem{\opone}$ 
from $\Values^n$ to $\ValuesE$.
Operators are essentially basic partial functions like destructors or
characteristic functions of predicates like $\ega$. 
The destructor $\head$ illustrates the purpose of $\erreur$ when it
satisfies $\sem{\head}(\vide) = \erreur$. 

A substitution $\sigma$ is a finite mapping from $\Var$ to $\Values$.
The application of a substitution $\sigma$ to an expression $\eqone$
is noted $\eqone\sigma$.


The language has a closure-based call-by-value semantics which is displayed in
Appendix~\ref{A}.
Given a substitution $\sigma$, the meaning of $\eqone\sigma \cbv \termfour$ is that is that $\eqone$ evaluates
to the value $\termfour$ of $\ValuesE$. If no rule is applicable, then
an error occurs, and $\eqone\sigma \cbv \erreur$.
A program $\p$ computes a partial function $\sem{\p} : \Values^n \to
\ValuesE$ defined by: For all $\termthree_i \in \Values,
\sem{\p}(\many{\termthree}{n}) = \termfour$ iff $
\p(\many{\termthree}{n}) \cbv \termfour$.  


\begin{example}[Division]\label{division}
Consider the following definitions that encode the division:
\begin{align*}
\minus(\varone,\vartwo) =\mathbf{Case\ } \varone,\vartwo \mathbf{ \ of\ }
\zero,\varthree &\to \zero \\
\suc(\varthree),\zero &\to \suc(\varthree) \\
\suc(\varfour),\suc(\varfive) &\to \minus(\varfour,\varfive)\\
\quo(\varone,\vartwo) =\mathbf{Case\ } \varone,\vartwo \mathbf{ \ of\ }
\zero,\suc(\varthree) &\to \zero\\
\suc(\varthree),\suc(\varfour) &\to \suc(\quo(\minus(\varthree,\varfour),\suc(\varfour)))
\end{align*}
Using the notation $\underline{n}$ for $\underbrace{\suc(\ldots \suc (\zero)\ldots)}_{n \text{ times } \suc}$, we have:
$$\sem{\quo}(\underline{n},\underline{m})=\underline{\lceil n/m \rceil} \text{ for $n,m>0$}$$
\end{example}

\section{Fraternities}\label{DP}
In this section, we define the notion of fraternity
based on dependency pairs, that Arts and Giesl~\cite{AG00} introduced 
to prove termination automatically. Fraternities will be used to tame the size of arguments of
recursive calls.
 
A \emph{context} is an expression  
$\conta{\many{\magique}{r}}$ containing one occurrence of each 
$\magique_i$. We suppose that the $\magique_i$'s are fresh variables
which are not in $\Voc$. The substitution of each $\magique_i$ by an expression $\eqtwo_i$ 
is noted $\conta{\many{\eqtwo}{r}}$.

\begin{definition}
Assume that  $\funon(\many{\varone}{n})=\eqone^{\funon}$ is a
definition of a program.
\emph{An expression $\eqtwo$ is activated} by $\funon(\many{\patone}{n})$
 where the $\patone_i$'s are patterns if there is a context with one hole $\conta{\magique}$ such that:
\begin{itemize}
\item If $\eqone^{\funon}$ is a compositional expression (that is
  with no case definition inside it), 
then 
$\eqone^{\funon}=\conta{\eqtwo}$. In this case, $\patone_{1} = \varone_{1} \ldots \patone_{n} = \varone_{n}$.

\item Otherwise, $
\black \eqone^{\funon} = \mathbf{Case\ } \many{e}{n} \mathbf{\ of\ } 
\overline{\pattwo_1} \rightarrow \eqone^1 
\ldots 
\overline{\pattwo_\ell} \rightarrow \eqone^\ell
$,
then there is a position $j$ such that
$\eqone^j=\conta{\eqtwo}$. 
In this case, $\patone_{1} = \pattwo_{j,1}\ldots \patone_{n} = \pattwo_{j,n}$ 
where $\overline{\pattwo_j} =  \pattwo_{j,1} \ldots \pattwo_{j,n}$. \black
\end{itemize}
\end{definition}

This definition is convenient in order to predict the
computational data flow involved. Indeed, an expression is activated by
$\funone(\many{\patone}{n})$ when  $\funone(\many{v}{n})$ is called 
and  each $v_i$ matches the corresponding pattern $\patone_{i}$.
An expression $\eqtwo$ activated by
$\funone(\many{\patone}{n})$ is \emph{maximal} if there is no context
$\conta{\magique}$, \black distinct from the empty context, \black such that 
$\conta{\eqtwo}$ is activated by $\funone(\many{\patone}{n})$.

\begin{definition}[Precedence]\label{prec}
 The notion of activated expression provides a precedence $\precF$ on function symbols. \black
Indeed, set $\funon \precF \funtw$ if there are $ \black \overline{e}$ and $\overline{\patone}$ such that 
$\funtw(\black \overline{e})$ is activated by $\funon(\overline{\patone})$.
Then, take the reflexive and transitive closure of $\precF$, that we also note
$\precF$. It is not difficult to establish that $\precF$ is a \black preorder\black.
Next, say that $\funon \equivF \funtw$ if $\funon \precF \funtw$ 
and inversely  $\funtw \precF \funon$.
Lastly, $\funon \sprecF \funtw$ if $\funon \precF \funtw$ 
and  $\funtw \precF \funon$ does not hold. Intuitively, $\funon \precF \funtw$  means that $\funon$ calls 
$\funtw$ in some executions. And  $\funon \equivF \funtw$ means
that  $\funon$ and $\funtw$ call themselves recursively.
\end{definition}

\begin{definition}[Fraternity]\label{fra}
 In a program $\p$, an expression $\conta{\funtw_{1}(\overline{\black e_1}),\ldots,\funtw_{r}(\overline{\black e_r})}$
activated by $\funon(\many{\patone}{n})$ is a fraternity if
\begin{enumerate}
\item $\conta{\funtw_{1}(\overline{\black e_1}),\ldots,\funtw_{r}(\overline{\black e_r})}$ is maximal
\item For each $i\in\{1,r\}$, $\funtw_{i} \equivF \funon$.
\item For every function symbol $\funthre$ that appears in the context $\conta{\many{\magique}{r}}$, we have $\funon \sprecF \funthre$.
\end{enumerate}
\end{definition}

A fraternity may correspond to a recursive call since it involves function symbols that are equivalent for the precedence $\precF$. 


\begin{example}
The program of example~\ref{division} admits two fraternities $\minus(\varfour,\varfive)$ and $\suc[\quo(\minus(\varthree,\varfour),\suc(\varfour))]$ which are respectively activated by $\minus(\suc(\varfour),\suc(\varfive))$ and $\quo(\suc(\varthree),\suc(\varfour))$.
\end{example}

%
%
%
%
%
%

\section{Sup-interpretations}
\begin{definition}[Partial assignment]\label{ass}
A partial assignment $\thetai$ is a partial mapping from the vocabulary
$\Voc$ which assigns a partial function $\thetai(b) : (\bR^+)^{n} \longmapsto \bR^+$ to each symbol $b$ in the domain of $\thetai$.
The domain of a partial assignment $\thetai$ is  noted $\dom(\thetai)$.
Because it is convenient, we shall always assume that partial
assignments that we consider, are defined on constructor and
operator symbols (i.e. $\Cns \cup \Op \subseteq \dom(\thetai)$). 

An assignment $\thetai$ is defined over an expression $\eqone$ if each
symbol of $\Cns \cup \Op \cup \Fct$ in $\eqone$ belongs to $\dom(\thetai)$. 
Suppose that the assignment $\thetai$ is defined over an expression $\eqone$ with $n$ variables. 
The partial assignment of $\eqone$ w.r.t. $\thetai$, that we note $\tstari(\eqone)$, is the canonical extension of
the assignment $\thetai$ and denotes a function from $(\bR^+)^n$ to $\bR^+$ defined as follows:
\begin{enumerate}
\item If $x_i$ is in $\Var$, let $\tstari(x_i)=X_i$ with $X_1,\ldots,X_n$ a sequence of new variables ranging over $\bR^+$.
%
%
\item  If $\overline{\eqone}$ is a sequence of $n$ expressions, then $\tstari(\overline{\eqone})=\max(\tstari(\eqone_1),\ldots,\tstari(\eqone_n))$
\item  If $\eqone$ is a $\mathbf{Case}$ expression of the shape $\mathbf{Case\ } \overline{\eqone} \mathbf{\ of\ } 
\overline{\patone_1} \rightarrow \eqone^1 
\ldots 
\overline{\patone_\ell} \rightarrow \eqone^\ell $, then $\tstari(\eqone)= \max(\tstari(\overline{\eqone}),\tstari(\eqone^1),\ldots, \tstari(\eqone^{\ell}))$
\item If $b$ is a 0-ary symbol or $b=\erreur$, then $\tstari(b)=\thetai(b)$.
\item If $b$ is a symbol of arity $n>0$ and $\many{\eqone}{n}$ are
  expressions,
  then we have
 $\tstari(b(\many{\eqone}{n}))  =\thetai(b)(\tstari(\eqone_{1}),\ldots,\tstari(\eqone_{n}))$
\end{enumerate}
\end{definition}

\begin{definition}[Additive assignments]\label{add}
 A partial assignment $\thetai$ is polynomial if for each symbol $b$
  of arity $n$ of $\dom(\thetai)$, $\thetai(b)$ is \textbf{bounded} by 
  a polynomial in $\bR^+[\many{X}{n}]$.
  An assignment of a constructor symbol $\conone$ is \emph{additive} if
$$  \thetai(\conone)(\many{X}{n})  = \sum_{i=1}^n X_i + \alpha_{\conone} \quad
  \alpha_{\conone} \geq 1$$

If the polynomial assignment of each constructor symbol is additive then the
assignment is additive. \textbf{Throughout the following paper we only consider additive assignments}.
\end{definition}

\begin{definition}
The size of an expression $\eqone$ is noted $\taille{\eqone}$ and defined by
$\taille{\eqone} = 0$ if $\eqone$ is a $0$-ary symbol or if $\eqone = \erreur$ and $\taille{b(\eqone_1,\ldots,\eqone_n)} = 1 + \sum_i \taille{\eqone_i}$ 
if $\eqone=b(\eqone_1,\ldots,\eqone_n)$ with $n>0$.
\end{definition}

\begin{lemma}\label{valeur}
Given an assignment $\thetai$, there is a constant $\alpha$ such that
for each value $v$ of $\ValuesE$, the following inequality is satisfied :
\begin{align*}
\taille{v} & \leq \tstari(v)  \leq  \alpha\taille{v} 
\end{align*}
\end{lemma}

\begin{definition}[Sup-interpretation]\label{SI} 
A sup-interpretation is a partial assignment $\theta$ which verifies the three
conditions below : 
\begin{enumerate}
\item \label{SI1}
 The assignment $\theta$ is weakly monotonic. That is, for each symbol
 $b \in \dom(\theta)$, the function $\theta(b)$ satisfies
\begin{align*}
\forall i=1,\ldots,n\ X_{i}\geq Y_{i} 
\Rightarrow 
\theta(b)(\many{X}{n}) \geq \theta(b)(\many{Y}{n})
\end{align*}  
\item \label{SI2}
For each $v \in \ValuesE$, 
\begin{align*} 
\tstar(v) & \geq \taille{v}
\end{align*}  
\item  \label{SI3}
For each symbol $b \in \dom(\theta)$ of arity $n$ 
and for each value $v_{1},\ldots,v_{n}$ of $\Values$,
if $\sem{b}(v_{1},\ldots,v_{n}) \in \ValuesE$,
then 
\begin{align*} 
\tstar(b(v_{1},\ldots,v_{n})) & \geq \tstar(\sem{b}(v_{1},\ldots,v_{n}))
\end{align*}
\end{enumerate}
We say that expression $\eqone$ admits a sup-interpretation $\theta$ if $\theta$ is defined over $\eqone$. 
The sup-interpretation of $\eqone$ wrt $\theta$ is $ \tstar(\eqone)$.
\end{definition} 

Intuitively, the sup-interpretation is a special program
interpretation. Instead of yielding the program denotation, a
sup-interpretation provides an upper bound on the output size of the function denoted by the program.
It is worth noticing that sup-interpretation is a complexity measure in the sense of Blum~\cite{B67}.

Given an expression $\eqone$, we define $\norme{\eqone}$ thus:
\begin{align*}
\norme{\eqone} & = 
\begin{cases}
\taille{\sem{\eqone}} & \text{if $\sem{\eqone} \in \ValuesE$} \\
0 & \text{otherwise} 
\end{cases}
\end{align*}

\begin{lemma}\label{sup-key}
Let $\eqone$ be an expression with no variable and which admits a sup-interpretation
$\theta$. If $\sem{\eqone} \in \ValuesE$ then have:
\begin{align*}
\norme{e}\leq \tstar(\sem{e}) \leq \tstar(e) 
\end{align*}
\end{lemma}

\begin{proof}
The proof is in~\cite{MP06}. \qed
%
\end{proof}




\begin{example}\label{Off}
Consider the program for exponential:
\begin{align*}
\expo(\varone) \ = \mathbf{Case\ } \varone \mathbf{\ of\ } 
\zero &\to \suc(\zero)\\
\suc(\vartwo) &\to \double(\expo(\vartwo))\\
\double(\varone)\ = \mathbf{Case\ } \varone \mathbf{\ of\ } 
\zero &\to \zero\\
\suc(\vartwo) &\rightarrow \suc(\suc(\double(\vartwo)))
\end{align*}
By taking $\theta(\suc)(X)=X+1$, $\theta(\double)(X)=2X$, we define a sup-interpretation of the function symbol $\double$. 
\end{example}

Now we are going to define the notion of weight which allows us to control the size of the arguments in
recursive calls.  A weight is an assignment having the subterm property but no longer
 giving a bound on the size of a value computed by a function. 

\begin{definition}[Weight]\label{wg} 
A weight $\weight$ is a partial assignment which ranges over $\Fct$. To a given function symbol $\funon$ of arity $n$ it assigns a total function $\omega_{\funon}$ from $(\bR^+)^n$ to $\bR^+$ which satisfies:
\begin{enumerate}
        \item $\weight_{\funon}$ is weakly monotonic.
$$
\forall i=1,\ldots,n,\ X_i\geq Y_i \Rightarrow \weight_{\funon}(\ldots,X_{i},\ldots) \geq \weight_{\funon}(\ldots,Y_{i},\ldots)
$$       
        \item $\weight_{\funon}$ has the subterm property 
$$
\forall i=1,\ldots,n,\ \forall X_i \in \bR^+ \ \weight_{\funon}(\ldots,X_{i},\ldots) \geq X_{i}
$$
\end{enumerate}
\end{definition} 

\begin{definition}[Call-tree]
\label{def:state}
A \emph{state} is a tuple $\nodeon$ where $\funon$ is a function
symbol of arity $n$ and $\valone_1, \ldots, \valone_n$ are values.  
Assume that $\eta_1 = \nodeon$ and $\eta_2 = \nodetw$ are two
states.
Assume also that  $\conta{\funtw(\many{\black e}{k})}$ is activated by
  $\funon(\many{\patone}{n})$.
A \emph{transition} is noted $\eta_1 \regleCT{} \eta_2$ and defined by:
\begin{enumerate}
\item There is a substitution $\sigma$ 
such that $\patone_i\sigma =  \valone_i$ for $i=1,\ldots,n$
\item and $\sem{\black e_j\sigma} = \valtwo_j$ for $j=1,\ldots,k$.
\end{enumerate}
We call such a graph a call-tree of $\funon$ over values $\valone_1, \ldots, \valone_n$ if $\nodeon$ is its root.
A state may be seen as a stack frame. A call-tree of root $\nodeon$ represents all the stack frames which will be pushed on the stack when we compute $\funon(\valone_1, \ldots, \valone_n)$.
\end{definition}

%
%
%
%
%

\section{Criterion to control space resources} 
\begin{definition}[Quasi-friendly]\label{QFC} 
A program $\p$ is \emph{quasi-friendly}  
iff there are a sup-interpretation $\theta$ and a weight $\omega$ such that for each fraternity of the shape
$\conta{\funtw_{1}(\overline{\black e_1}),\ldots,\funtw_{r}(\overline{\black e_r})}$, activated by $\funon(\many{\patone}{n})$, we have:
\begin{enumerate}
\item $\weight_{\funon}(\tstar(\patone_{1}),\ldots,\tstar(\patone_{n})) \geq \max_{i=1..r}(\weight_{\funtw_{i}}(\tstar (\overline{e_{i}})))$\\
\item $\weight_{\funon}(\tstar(\patone_{1}),\ldots,\tstar(\patone_{n})) \geq \tstar(\Conb)[\weight_{\funtw_{1}}(\tstar (\overline{e_{1}})),\ldots,\weight_{\funtw_{r}}(\tstar (\overline{e_{r}}))]$
\end{enumerate}
\end{definition}

Notice that nested fraternities (i.e. a fraternity $\eqtwo$ containing another fraternity inside it) are not of real interest for this criterion. In fact, consider for example the following nested fraternity $\funone(\varone)=\funone(\funone(\varone))$. In the quasi-friendly criterion, one need to guess a weight and a sup-interpretation for the function symbol $\funone$, so that, the criterion becomes useless.
However this is not a severe drawback since such programs are not that natural in a programming perspective and
either they have to be really restricted or they rapidly generate complex functions like the Ackermann one.

Since $\tstar$ has no subterm property, conditions 1 and 2 are independent and useful in order to control the size of the values added by recursive calls. An example showing this independence is given in appendix~\ref{E}.

\begin{theorem}\label{THQFC}
Assume that $\p$ is a quasi-friendly program, then for each function symbol $\funon$ of $\p$ there is a polynomial
  $P$  such that for every value $v_{1},\ldots,v_{n}$,
\begin{align*}
\norme{\funon(v_{1},\ldots,v_{n})}
    & \leq P(\max(\taille{v_{1}},...,\taille{v_{n}}))
\end{align*}
\end{theorem}
\begin{proof}
The proof can be found in appendix~\ref{C}.
\qed
\end{proof}

\begin{example}
The program of example~\ref{division} is quasi-friendly. Taking:
$$\begin{array}{rl p{1cm} | p{1cm} rl}
\theta(\suc)(X)&=X+1 &&& \theta(\zero) &=0 \\
\theta(\minus)(X,Y) & =X &&& \weight_{\minus}(X,Y) &= \max(X,Y)\\
\weight_{\quo}(X,Y) &= X+Y &&& &
\end{array}$$
We check the conditions for the fraternity defined by $\quo$:
\begin{align*}
\weight_{\quo}(\tstar(\suc(z)),\tstar(\suc(u))) &=U+Z+2\\
&\geq Z+U+1\\
&=\weight_{\quo}(\tstar(\minus(z,u)),\tstar(\suc(u))) &&\text{(Condition 1)}\\
\weight_{\quo}(\tstar(\suc(z)),\tstar(\suc(u))) &=U+Z+2\\
&\geq Z+U+2\\
&=\tstar(\suc)(\weight_{\quo}(\tstar(\minus(z,u)),\tstar(\suc(u))))&&\text{(Condition 2)}
\end{align*}
\end{example}

\begin{example}
The program of example~\ref{Off} is not quasi-friendly. Indeed since the sup-interpretation of $\double$ is greater than $2X$. One has to find a polynomial weight $\weight_{\exp}$ such that:
$$\weight_{\exp}(X+1) \geq \theta(\double)(\weight_{\exp}(X)) \geq 2\weight_{\exp}(X) $$
which is impossible.
\end{example}

\begin{theorem}\label{THQFCT}
Assume that $\p$ is a quasi-friendly program.
For each function symbol $\funon$ of $\p$ there is a polynomial
  $R$ such that for every node $\nodetwobb$ of the call-tree of root $\nodeonebb$, 
\begin{align*}
\max_{j=1..m}(\taille{u_j})
    & \leq R(\max(\taille{v_{1}},...,\taille{v_{n}}))
\end{align*}
even if $\funon(v_{1},\ldots,v_{n})$ is not terminating.
\end{theorem}
\begin{proof}
The proof relies on theorem~\ref{THQFC} and is essentially the same than the one in~\cite{MP06}.\qed
\end{proof}

In the paper~\cite{MP06}, a first criterion, called friendly criterion, was developed in order to bound the stack frame size during the execution of a program.
However, as mentioned in the conclusion of~\cite{MP06}, this criterion was suffering from a lack because of a too restrictive condition on the contexts. Indeed, the sup-interpretations of the contexts were forced to be $\max$ functions forbidding, for example, recursion over tree data structure as in the example of Appendix~\ref{D}.
Thus, from practical experience, the quasi-friendly criterion captures more algorithms than the friendly criterion.

\section{Comparison with quasi-interpretations}


\begin{definition}\label{QI}
A quasi-interpretation is a total (i.e. defined for every symbol of the program) additive assignment $\interp{-}$ monotonic and having the subterm property (i.e. $ \text{ For all symbol } \funon \text{ of arity } n, \forall i \in \left\{1,n\right\}, \interp{\funon}(\ldots,X_i,\ldots) \geq X_i$) such that for every maximal expression $e$ activated by $\funon(\many{p}{n})$ we have:
$$\interp{\funon(\many{p}{n})} \geq \interp{e}$$ where the assignment $\interp{-}$ is extended canonically to terms by $$\interp{\funtw(\many{e}{n})}=\interp{\funtw}(\interp{e_1},\ldots,\interp{e_n})$$
\end{definition}

As demonstrated in~\cite{BMM01,BMM05b,MM00}, quasi-interpretations have the following property:

\begin{proposition}
Given a program $\p$  which admits a quasi-interpretation $\interp{-}$, for each function symbol $\funon$ of $\p$ and any $v,\many{v}{n} \in \Values$,
\begin{align*}
\interp{\funon}(\interp{v_1},\ldots,\interp{v_n}) &\geq \interp{\sem{\funon}(\many{v}{n})}\\
\interp{v} &\geq \taille{v}
\end{align*}
\end{proposition}

\begin{theorem}\label{2}
Every quasi-interpretation is a sup-interpretation.
\end{theorem}
\begin{proof}
By previous proposition, conditions 2 and 3 of Definition~\ref{SI} hold. By Definition~\ref{QI}, a quasi-interpretation is monotonic, so that condition 1 of Definition~\ref{SI} holds.
\qed
\end{proof}
A very interesting consequence of this Theorem concerns the sup-interpretation synthesis problem. The synthesis problem consists in finding a sup-interpretation for a given program. It was introduced by Amadio in~\cite{Amadio03} for quasi-interpretations. This problem is very relevant in a perspective of automating the complexity analysis of programs. However the synthesis of quasi-interpretation is a very tricky problem which is undecidable in general. However Amadio showed~\cite{Amadio03} that some rich classes of quasi-interpretation are in NP and in~\cite{BMMP05}, it was demonstrated that the quasi-interpretation synthesis with bounded polynomials over reals is decidable. Consequently, we get some heuristics for the synthesis of sup-interpretation in $\textbf{Max-Poly}$, the set of functions defined to be constant functions, projections, $\max$, $+$, $\times$ and closed by composition: Given a program $\p$, we try to find a quasi-interpretation for this program, and, by previous Theorem, we know that it is a sup-interpretation.

\begin{theorem}\label{1}
Every program that admits a quasi-interpretation is quasi-friendly.
\end{theorem}
\begin{proof}
By previous theorem every quasi-interpretation defines a sup-interpreta\-tion. Moreover every quasi-interpretation is a weight. 
\end{proof}

\begin{proposition}
There exist quasi-friendly programs that do not have any quasi-interpretation.
\end{proposition}
\begin{proof}
Program of example~\ref{division} is quasi-friendly but does not admit any quasi-interpretation. In fact, suppose that it admits an additive quasi-interpretation $q$. For the last definition, we have:
\begin{align*}
\interp{\quo(\suc(\varfive),\suc(\varfour))} &=\interp{\quo}(U+k,V+k) &&\text{For some constant $k$}\\
&\geq \interp{\suc(\quo(\minus(\varfive,\varfour),\suc(\varfour)))} &&\text{By Dfn of $\interp{-}$}\\
&\geq k+\interp{\quo}(\max(U,V),U+k) \\
&> \interp{\quo}(U+k,V+k) &&\text{for } V \geq U+1 
\end{align*}
Consequently, we obtain a contradiction and $\quo$ does not admit any quasi-interpre\-tation. 
\qed
\end{proof}

In~\cite{BMM01,BMM05b,MM00}, some characterizations of the functions computable in polynomial time and polynomial space were given. Theorems~\ref{THQFC} and~\ref{2} allow to adapt these results to the sup-interpretations.

Given a precedence (quasi-order) $\precFC$ on $\Constructors \cup \Fct$.
Define the equivalence relation $\equivFC$ as $\funone \equivFC \funtwo$
iff $\funone \precFC \funtwo$ and 
$\funtwo \precFC \funone$. 
We associate to each function symbol $\funone$ a status $st(\funone)$
in $\left\{p,l\right\}$ and satisfying if $\funone \equivFC \funtwo$
then $st(\funone)=st(\funtwo)$. 
The status indicates how to compare recursive calls. 

\begin{definition}\label{def:Muliset-ordering}
The product extension $\infMulSt$ and the lexicographic extension $\prec^l$ of $\infSt$ over sequences are defined by:
\begin{itemize}
\item $ (\many{m}{k}) \infMulSt (\many{n}{k})$ if and only if (i)
  $\forall i \leq k, m_i \infEq n_{i}$ and (ii) $\exists j \leq k$
  such that $m_j \infSt n_{j}$.
\item $ (\many{m}{k})\prec^l (\many{n}{l})$ if and only if $\exists j$ such that $\forall i<j,\  m_i \preceq n_i$ and $m_j \prec n_j$ 
\end{itemize}
\end{definition}

\begin{definition}
Given a precedence $\precFC$ and a status $st$, 
we define the recursive path ordering $\xpoSt$ as follows:
\begin{gather*}
\ninfer{\termtwo \xpoEq \termone_i}{\termtwo \xpoSt \funone(\ldots,
\termone_i, \ldots)}{} \qquad
\ninfer{ \forall i\ \termtwo_i \xpoSt \funone(\many{\termone}{n}) \qquad
g \precFC \funone}{g(\many{\termtwo}{m}) \xpoSt
\funone(\many{\termone}{n})}{}\\[5mm]
\infer{(\many{\termtwo}{n}) \xpoXSt{st(\funone)} (\many{\termone}{n}) \qquad
\funone \equivFC \funtwo \qquad \forall i\ \termtwo_i \xpoSt
\funone(\many{\termone}{n})}{\funtwo(\many{\termtwo}{n}) \xpoSt
\funone(\many{\termone}{n})}
\end{gather*}
The $\mathbf{Case} \ldots \mathbf{of} \ldots \to$ (and the symbol $=$ in a definition without $\mathbf{Case}$) expressions induce a rewrite relation noted $\to$.
A program is ordered by $\xpoSt$ if there are a precedence
$\precEqFS$ and a status $st$ such that for each rule $l \to r$ of the rewrite relation, the
inequality $r \xpoSt l$ holds. 
\end{definition} 

\begin{theorem} \quad
\begin{itemize}
\item The set of functions computed by quasi-friendly programs admitting an additive sup-interpre\-tation and ordered by $\xpoSt$ where each function symbol has a product status  is exactly the set of functions computable in polynomial time.
\item The set of functions computed by quasi-friendly programs admitting an additive sup-interpre\-tation and ordered by $\xpoSt$ is exactly the set of functions computable in polynomial space.
\end{itemize}
\end{theorem}
\begin{proof}
We give here the main ingredients of the proof. The main idea of the proof is fully written in~\cite{BMM05b}. Due to the $\xpoSt$ ordering with product status, any recursive subcall of some $\funone(\many{v}{n})$, with $\funone$ function symbol and $v_i$ constructor terms, will be done on subterms of the $v_i$. A consequence of Theorem~\ref{THQFC} is that any other subcalls will be done on arguments of polynomial size. So one may use a memoization technique a la Jones~\cite{JonesCC}, which leads us to define a call-by-value interpreter with cache in Appendixé~\ref{B}.
\qed
\end{proof}

\bibliography{bib}
\bibliographystyle{plain}
\newpage
\appendix
\section{Call-by-value semantics}\label{A}\quad
\begin{figure}[h]
\hrule
\begin{gather*}
\ninfer 
{ 
  \termone_1  \cbv \termfour_1 \ldots \termone_n  \cbv \termfour_n} 
{  \conone(\many{\termone}{n}) \cbv  
  \conone(\many{\termfour}{n})}  
{\conone \in \Cns \text{\ and\ } \forall i,\termfour_i\not=\erreur}
\\[0.5cm] 
\ninfer 
{
  \termone_1  \cbv \termfour_1 \ldots \termone_n  \cbv \termfour_n} 
{  \opone(\many{\termone}{n}) \cbv  
  \sem{\opone}(\many{\termfour}{n})}  
{\opone \in \Op \text{\ and\ }  \forall i,\termfour_i\not=\erreur}
\\[0.5cm] 
\ninfer 
{
  \eqone  \cbv u  
  \quad \exists \sigma,\ i \ :\ \patone_i\sigma=u
  \quad  \eqone_i\sigma \cbv \termfour}
{  \CasPlat{e}\cbv  \termfour  }
{\text{$\mathbf{Case}$}  \text{\ and\ } u\not=\erreur }
\\[0.5cm] 
\ninfer { \eqone_1  \cbv \termfour_1 \ldots \eqone_n  \cbv
  \termfour_n \quad
\funone(\many{\varone}{n})= \eqone^{\funone} \quad 
 \eqone^{\funone}\sigma \cbv \termfour} 
{\funone(\many{\eqone}{n})  \cbv \termfour} {\text{where\ } \sigma(\varone_i)=\termfour_i} \not=\erreur \text{\ and\ } \termfour\not=\erreur
\end{gather*}
\caption{Call by value semantics of ground expressions wrt a program $\p$} 
\label{fig:cbv}
\hrule
\end{figure}

\section{Example}\label{E}
The following non-terminating program illustrates that conditions 1 and 2 of the quasi-friendly criterion are independent.
\begin{align*}
\half(t) = \mathbf{Case\ } t \mathbf{\ of\ } \suc(\suc(x)) &\to \suc(\half(x))\\
\suc(\zero) &\to \zero\\
\zero &\to \zero\\
\funone(x)=\half(\funone(\double(x))&
\end{align*}
where $\double$ is the function of example~\ref{Off}. The arguments of $\funone$ computed by the recursive calls are unbounded. However by taking $\theta(\half)(X)=X/2$, $\theta(\double)(X)=2X$ and $\weight_{\funone}(X)=X$, we can check that the Condition 2 of the quasi-friendly criterion is satisfied, even if Condition 1 is not.

\section{Proof of Theorem~\ref{THQFC}}\label{C}
We start by showing the following lemma:
\begin{lemma}\label{borne}
If a locally friendly program has a call-tree containing a branch of the shape $\nodeon \regleCT{*} \nodetw$ with $\funon \equivF \funtw$ then:
$$\weight_{\funon}(\ttmany{u}{n}) \geq \weight_{\funtw}(\ttmany{v}{k})$$
\end{lemma}
\begin{proof}
We show it by induction on the number $n$ of states in the branch:
\begin{itemize}
\item If $n=1$, $\nodeone \regleCT{} \nodetwo$ then there is a definition with a fraternity of the shape $\funone(\many{x}{n}) =\mathbf{Case\ } \many{x}{n} \mathbf{\ of\ } \many{p}{n} \to \conta{\funtwo(\many{e}{k})}$ with $\funone \equivF \funtwo$ and a substitution $\sigma$ such that $p_i\sigma = u_i$ and $\sem{e_j\sigma}=v_j$. Applying the Condition 1 of the quasi-friendly criterion, we obtain:
$$\weight_{\funone}(\ttmany{u}{n}) \geq \weight_{\funtwo}(\tmany{e}{k})\\
\geq \weight_{\funtwo}(\ttmany{v}{k})$$ By monotonicity of weights and by definition of sup-interpretations.
\item Now suppose by induction hypothesis that if $\nodeone \regleCT{k} \nodetwo$ with $\funone \equivF \funtwo$ and $k\leq n$, we have $$\weight_{\funone}(\ttmany{u}{n}) \geq \weight_{\funtwo}(\ttmany{v}{k})\quad (I.H.)$$
And consider the following branch of length $n+1$:
$$\nodeone \regleCT{n} \nodetwo \regleCT{} \nodethree$$
with $\funthree \equivF \funone$.
Then as in the base case, we can derive 
$$\weight_{\funtwo}(\ttmany{v}{k}) \geq \weight_{\funthree}(\ttmany{v'}{l})$$
and combine it with the Induction Hypothesis to obtain:
$$\weight_{\funone}(\ttmany{u}{n}) \geq \weight_{\funthree}(\ttmany{v'}{l})$$
\end{itemize}
\qed
\end{proof}
\
\setcounter{theorem}{0}
\begin{theorem}
Assume that $\p$ is a quasi-friendly program.
For each function symbol $\funon$ of $\p$ there is a polynomial
  $P$  such that for every value $v_{1},\ldots,v_{n}$,
\begin{align*}
\norme{\funon(v_{1},\ldots,v_{n})}
    & \leq P(\max(\taille{v_{1}},...,\taille{v_{n}}))
\end{align*}
\end{theorem}

\begin{proof}
Suppose that we have a program $\p$ and a function symbol $\funone \in \Fct$ and $\many{v}{n} \in \Values$ such that $\sem{\funone}(\many{v}{n})$ is defined (i.e. the function computation terminates on inputs $\many{v}{n}$). We are going to show the previous result by an induction on the precedence $\precF$.
\begin{itemize}
\item If $\funone$ is defined without function symbols (i.e. $\funone$ is strictly smaller than any other function symbol for $\precF$), then a definition of the shape $\funone(\many{x}{n}) = e$ with $e \in \CVterms$ is applied. We define $P_{\funone}(X)=\taille{e}$ with the size of a variable $y$ being defined by $\taille{y}=X$. Taking a substitution $\sigma$ such that $p_i \sigma = v_i$, we can check easily that $$P_{\funone}(\max_{i=1..n}\taille{v_i})= \taille{e[X:=\max_{i=1..n}\taille{v_i}]}\geq \taille{e \sigma}= \norme{\funone(\many{v}{n})}$$ where $\taille{e[X:=\taille{v}]}$ denotes the substitution of the variable $X$ by the value $\taille{v}$ in the function $\taille{e}$.
\item Now, if the function symbol $\funone$ is defined without fraternities, then we have definitions of this shape $\funone(\many{x}{n}) =\mathbf{Case\ } \many{x}{n} \mathbf{\ of\ } \many{p}{n} \to e$ with for all function symbol $\funtwo \in e, \funone \sprecF \funtwo$. We suppose by induction hypothesis that we have already defined a polynomial upper bound on the function symbols $\funtwo$. Moreover, for every constructor symbol $\conone \in e$ of arity $n$, we define $P_{\conone}(X)=nX+1$, which represents a polynomial upper bound on its computation (i.e. the constructor symbol keeps its arguments and adds 1 to the global size). Finally, if $e=\funthree(\many{e}{m})$, we define inductively a polynomial upper bound on the size of the computation of $e$ by $P_e(X)=P_{\funthree}(\max_{i=1..m}P_{e_i}(X))$. By definition of such a polynomial, we know that $P_e(\max_{i=1..n}\taille{v_i}) \geq \norme{f(\many{v}{n})}$.
\item Now, suppose that the function symbol is defined with some definitions leading to fraternities and some definitions similar to the one of the previous case (i.e. definitions which are not recursive). First, we build a polynomial $P_{\funone \sprecF}$, as in the previous case, for these latter definitions. Notice also that since we know, by hypothesis, that the computation is terminating, every recursive call will be ended by such definitions. However it can be ended by such a definition for some other equivalent function symbol. Thus for each $\funtwo \equivF \funone$, we also define $P_{\funtwo \sprecF}$ and finally, we define a new polynomial $Q_{\funone}(X)=\max_{\funtwo \equivF \funone}(P_{\funtwo \sprecF}(X))$. Intuitively, this polynomial is an upper bound on the size of every value computed by a definition which will leave a dependency pair cycle in Arts and Giesl's work. Now, combining condition 2 of Definition~\ref{QFC} and lemma~\ref{borne}, we know that if for some values $\many{v}{n}$, $\funone(\many{v}{n}) \reftransto \conta{\funtwo_{1}(\overline{u_1}),\ldots,\funtwo_{r}(\overline{u_l})}$ with $\funtwo_1 \equivF \ldots \equivF \funtwo_l \equivF \funone$ and $\to$ the rewrite relation induced by the definitions of the program, then:
\begin{align}
\weight_{\funone}(\ttmany{v}{n}) &\geq \tstar_{\overline{v}}(\mathsf{C})[\weight_{\funtwo_{1}}(\tstar_{\overline{v}}(\overline{u_1})),\ldots,\weight_{\funtwo_{l}}(\tstar_{\overline{v}}(\overline{u_l}))]
\end{align}
where the notation $\tstar_{\overline{v}}(e)$ means that the sup-interpretation of $e$ may depend on $\overline{v}=\many{v}{n}$.

This result holds particularly in the case where the $\funtwo_{i}(\overline{u_i})$ correspond to function calls that will leave the recursive call (i.e. function symbols that call function symbols strictly smaller for the precedence). Since we are considering defined values (i.e. evaluations that terminate), such calls exist. By condition 2 of Definition~\ref{SI}, we know that $\tstar(\overline{u_i}) \geq \taille{\overline{u_i}}$. By subterm property of weights, we obtain $\weight_{\funtwo_{i}}(\tstar(\overline{u_i})) \geq  \max{\taille{\overline{u_i}}}$ and since $Q_{\funone}$ is monotone (by construction) $Q_{\funone}(\weight_{\funtwo_{i}}(\tstar(\overline{u_i}))) \geq Q_{\funone}(\max{\taille{\overline{u_i}}})$. Now, since sup-interpretations represent an upper bound on the values computed by the functions, if we have $\conta{\funtwo_{1}(\overline{ u_1}),\ldots,\funtwo_{l}(\overline{u_r})} \cbv \sem{\funone}(\many{v}{n})$ then by monotonicity of sup-interpretations, weights and $Q_{\funone}$:
\begin{align*}
\tstar_{\overline{v}}(\mathsf{C})[Q_{\funone}(\weight_{\funtwo_{1}}( \tstar_{\overline{v}}(\overline{u_1}))),\ldots,Q_{\funone}(\weight_{\funtwo_{l}}(\tstar_{\overline{v}}(\overline{u_l})))] &\geq \\
\tstar_{\overline{v}}(\mathsf{C})[Q_{\funone}(\max{\taille{\overline{u_1}}}),\ldots,Q_{\funone}(\max{\taille{\overline{u_l}}}))]&\geq \norme{\funone(\many{v}{n})}
\end{align*}
It remains to show that the left-hand side of this inequality is bounded polynomially in the size of the inputs. Inequality (1), implies that $\tstar_{\overline{v}}(\mathsf{C})[\many{\magique}{l}]$ is polynomial in $\magique_j$ whenever $\weight_{\funtwo_{j}}(\tstar_{\overline{v}}(\overline{u_j}))$ depends on $\overline{v}$ (Else we obtain a contradiction since $\weight_{\funone}(\ttmany{v}{n})$ is polynomial in the $\ttmany{v}{n}$. Moreover, if $\weight_{\funtwo_{j}}(\tstar_{\overline{v}}(\overline{u_j}))$ does not depend on $\overline{v}$ then it is constant. By lemma~\ref{borne} and by monotonicity of $Q_{\funone}$, $Q_{\funone}(\weight_{\funtwo_{j}}(\tstar_{\overline{v}}(\overline{u_j})))$ is bounded by $Q_{\funone}(\weight_{\funone}(\tstar(\overline{v})))$.

Finally, the ring of polynomials being closed by composition, we know that $\norme{\funone(\many{v}{n})}$ is polynomially bounded in the $\ttmany{v}{n}$. Since the considered sup-interpretations are additive, we have by lemma~\ref{valeur} that $\tstar(v)  \leq  \alpha\taille{v} $ for some constant $\alpha$. Consequently, $\norme{\funone(\many{v}{n})}$ is also bounded by a polynomial in $\taillemany{v}{n}$ which is independent from the inputs.
\end{itemize}
\qed
\end{proof}

\section{Example}\label{D}
The following example illustrates that the quasi-friendly criterion captures, in practice, more algorithms than the friendly criterion of~\cite{MP06}. In fact, contrary to this latter criterion, the quasi-friendly criterion captures algorithms over trees (where the tree algebra is generated by the binary constructor symbol $\conone$ for nodes and the unary constructor symbol $\tap$ for leaves).
\begin{align*}
\funone(s,t)= \mathbf{Case\ } s,t \mathbf{\ of\ }  \conone(x,y),\conone(x',y') &\to \conone(\funone(x,y),\funone(x',y'))\\
\conone(x,y),\tap(u) &\to \tap(u)\\
\tap (u),\conone(x,y) &\to \tap (u)\\
\tap (u),\tap (v) &\to \quo(u,v)
\end{align*}

If the leaves of s and t are the words $\many{u}{n}$ and $\many{v}{n}$, then $\funone$ computes the tree whose leaves form the word $\quo(u_1,v_2),\ldots,\quo(u_n,v_n)$ with $\quo$ the division function described in example~\ref{division}.
Taking $\weight_{\funone}(X,Y)=X+Y$, $\theta(\tap)(X)=X+1$, $\theta(\quo)(X,Y)=X$ and $\theta(\conone)(X,Y)=X+Y+1$ we can show easily that it is quasi-friendly. 
\begin{align*}
\weight_{\funone}(\tstar(\conone(x,y)),\tstar(\conone(x',y'))&=X+Y+X'+Y'+2 \\
&> \max(X+Y,X'+Y')\\
&=\max(\weight_{\funone}(\tstar(x),\tstar(y),\weight_{\funone}(\tstar(x'),\tstar(y')) &
&\text{(Cnd 1)}\\
\weight_{\funone}(\tstar(\conone(x,y)),\tstar(\conone(x',y'))&=X+Y+X'+Y'+2  \\
&>X+Y+X'+Y'+1\\
&= \theta(\conone)(\weight_{\funone}(\tstar(x),\tstar(y)),\weight_{\funone}(\tstar(x'),\tstar(y'))) &&\text{(Cnd 2)}
\end{align*}

\newpage

\section{Interpreter with cache}\label{B}
\begin{figure*}
\begin{scriptsize}
\begin{gather*}
\ninfer { \substi{\varone} = \termfour} {\Equations, \subst \vdash \langle C, \varone \rangle \to \langle C, \termfour \rangle}
	 {(Variable)} \qquad
\ninfer {\conone \in \Constructors \quad \Equations, \subst \vdash \langle C_{i-1}, \termone_i \rangle \to \langle C_i, \termfour_i \rangle} {\Equations, \subst \vdash \langle C_0,
\conone(\many{\termone}{n}) \rangle \to \langle C_n,
\conone(\many{\termfour}{n}) \rangle} {(Cons)}\\ \\
\ninfer {\funone \in \Fct \quad \Equations, \subst \vdash \langle C_{i-1}, \termone_i \rangle \to \langle C_i, \termfour_i \rangle \quad (\funone(\many{\termfour}{n}),
\termfour) \in C_n} {\Equations, \subst \vdash \langle C_0,
\funone(\many{\termone}{n}) \rangle \to \langle C_n, \termfour \rangle} {(Cache \ reading)}\\ \\
\ninfer {\Equations, \subst
\vdash \langle C_{i-1}, \termone_i \rangle \to \langle C_i, \termfour_i \rangle \quad
\funone(\many{\patone}{n}) \to r \in \Equations \quad \patone_i \subst' = \termfour_i \quad \Equations, \subst' \vdash \langle C_n, r \rangle \to \langle C,\termfour \rangle} {\Equations, \subst \vdash \langle C_0,\funone(\many{\termone}{n}) \rangle \to \langle C \union (\funone(\many{\termfour}{n}),\termfour), \termfour \rangle} {(Push)} 
\end{gather*} 
\end{scriptsize} 
\caption{Evaluation of a rewriting system with memoization of intermediate evaluations} \label{fig:interp-cache} 
\end{figure*}
\end{document}

%% file: macroSI.tex
\newcommand{\p}{\mbox{\textbf{p}}}

\edef\termone{t}
\edef\termtwo{u}
\edef\termthree{v}
\edef\termfour{w}

\input prooftree
\newcommand{\infer}[2]
     {\prooftree
          #1 
          \justifies #2
      \endprooftree}

\newcommand{\ninfer}[3]
     {\prooftree
          #1 
          \justifies #2
          \using #3
      \endprooftree}
\newcommand{\some}[3]{#1_{#2}, \cdots, #1_{#3}}
\newcommand{\many}[2]{\some{#1}{1}{#2}}
\newcommand{\tsome}[3]{\tstar(#1_{#2}\sigma), \cdots, \tstar(#1_{#3}\sigma)}
\newcommand{\tmany}[2]{\tsome{#1}{1}{#2}}
\newcommand{\ttsome}[3]{\tstar(#1_{#2}), \cdots, \tstar(#1_{#3})}
\newcommand{\ttmany}[2]{\ttsome{#1}{1}{#2}}

\newcommand{\taille}[1]{|#1|}

\newcommand{\bR}{\mathbb{R}}

\newcommand{\Fct}{\textit{Fct}}
\newcommand{\Var}{\textit{Var}}
\newcommand{\Op}{\textit{Op}}
\newcommand{\Cns}{\textit{Cns}}
\newcommand{\Voc}{\Sigma}

\newcommand{\tstar}{\theta^*}

\newcommand{\dom}{\text{dom}}

\newcommand{\defone}{\mathit{def}}

\newcommand{\Equations}{\mathcal{E}}
\newcommand{\eqone}{e}
\newcommand{\eqtwo}{d}

\newcommand{\opone}{\mathbf{op}}

\newcommand{\CasPlat}[1]{
\mathbf{Case\ }  #1\mathbf{\ of\ } 
\patone_1 \rightarrow #1_1 
\ldots 
\patone_\ell \rightarrow #1_\ell 
}

\newcommand{\Values}{\mathit{Values}}
\newcommand{\ValuesE}{\mathit{Values}^{*}}

\newcommand{\erreur}{\mathbf{Err}}

\newcommand{\conone}{\mathbf{c}}

\newcommand{\edge}{\mbox{\textbf{edge}}}
\newcommand{\Variables}{\mathcal{X}}
\newcommand{\varone}{x}
\newcommand{\vartwo}{y}
\newcommand{\varthree}{z}
\newcommand{\varfour}{u}
\newcommand{\varfive}{v}

\newcommand{\patone}{p}
\newcommand{\pattwo}{q}

\newcommand{\funone}{\texttt{f}}
\newcommand{\funtwo}{\texttt{g}}
\newcommand{\funthree}{\texttt{h}}

\newcommand{\double}{\texttt{double}}
\newcommand{\expo}{\texttt{exp}}

\newcommand{\add}{\texttt{add}}

\newcommand{\half}{\texttt{half}}

\newcommand{\ou}{\mbox{\texttt{or}}}
\newcommand{\union}{\mbox{\texttt{union}}}
\newcommand{\equi}{\mbox{\texttt{equi}}}
\newcommand{\reach}{\mbox{\texttt{reach}}}
\newcommand{\si}{\mbox{\texttt{if}}}
\newcommand{\les}{\mbox{\texttt{le}}}
\newcommand{\app}{\mbox{\texttt{app}}}
\newcommand{\low}{\mbox{\texttt{low}}}
\newcommand{\high}{\mbox{\texttt{high}}}
\newcommand{\qs}{\mbox{\texttt{qs}}}

\newcommand{\minus}{\mbox{\texttt{minus}}}
\newcommand{\quo}{\mbox{\texttt{q}}}
\newcommand{\remove}{\mbox{\texttt{remove}}}
\newcommand{\elim}{\mbox{\texttt{elim}}}

\newcommand{\adds}{\mbox{\texttt{addstream}}}
\newcommand{\valone}{u}
\newcommand{\valtwo}{v}

\newcommand{\zero}{\textbf{0}}
\newcommand{\true}{\mbox{\textbf{True}}}
\newcommand{\false}{\mbox{\textbf{False}}}
\newcommand{\nil}{\textbf{nil}}
\newcommand{\suc}{\textbf{S}}

\newcommand{\vide}{\textbf{nil}}

\newcommand{\head}{\textbf{hd}}

\newcommand{\sem}[1]{\llbracket #1 \rrbracket}
\newcommand{\norme}[1]{\lVert#1\rVert}

\newcommand{\cbv}{\downarrow}

\def\cont#1#2{\mathsf{#1}[#2]}

\def\conta#1{\cont{C}{#1}}

\def\magique{\diamond}

\newcommand{\precF}{\geq_{\mathcal{F}}}
\newcommand{\sprecF}{>_{\mathcal{F}}}
\newcommand{\equivF}{\approx_{\mathcal{F}}}

\newcommand{\weight}{\omega}

\newcommand{\nodeone}{\langle \funone, \many{\valone}{n} \rangle}

\newcommand{\nodetwo}{\langle \funtwo, \many{\valtwo}{k} \rangle}

\newcommand{\nodethree}{\langle \funthree, \many{\valtwo'}{l} \rangle}

\newcommand{\regleCT}[1]{\stackrel{\mbox{\scriptsize $#1$}}{\leadsto}}

\newcommand{\ega}{\textbf{=}}
\newcommand{\funon}{\emph{\texttt{f}}}
\newcommand{\funtw}{\emph{\texttt{g}}}
\newcommand{\Const}{\Terms{\Cns}}
\newcommand{\funthre}{\emph{\texttt{h}}}
\newcommand{\thetai}{I}
\newcommand{\tstari}{\thetai^*}

\newcommand{\Terms}[1]{\mathcal{T}(#1)}
\newcommand{\nodeon}{\langle \emph{\funone}, \many{\valone}{n} \rangle}
\newcommand{\nodetw}{\langle \emph{\funtwo}, \many{\valtwo}{k} \rangle}
\newcommand{\Conb}{\mathsf{C}}
\newcommand{\nodeonebb}{\langle\emph{\funone}, \many{\valtwo}{n} \rangle}
\newcommand{\nodetwobb}{\langle \emph{\funtwo}, \many{\valone}{m} \rangle}
\newcommand{\interp}[1]{\llparenthesis #1 \rrparenthesis}
\newcommand{\precFC}{\geq_{\textit{Fct}\ \cup \textit{Cns}}}

\newcommand{\equivFC}{\approx_{\textit{Fct}\ \cup \textit{Cns}}}
\newcommand{\Constructors}{\textit{Cns}}
\edef\infMulSt{\prec^p}
\edef\infSt{\prec}
\edef\infEq{\preceq}
\edef\xpoSt{\prec_{rpo}}
\edef\xpoEq{\preceq_{rpo}}
\newcommand{\xpoXSt}[1]{\prec^{#1}_{rpo}}

\newcommand{\precEqFS}{\preceq_{\Functions}}

\newcommand{\Functions}{\textit{Fct}}
\newcommand{\CVterms}{\Terms{\Constructors \cup \Variables}}
\def\reftransto{\mbox{${\stackrel{\tiny *\;}{\To}}$}}
\newcommand{\taillesome}[3]{\taille{#1_{#2}}, \cdots, \taille{#1_{#3}}}
\newcommand{\taillemany}[2]{\taillesome{#1}{1}{#2}}

\def\To{\to}
\newcommand{\substi}[1]{{\sigma(#1)}}
\newcommand{\subst}{\sigma}
\newcommand{\tap}{\mathbf{tip}}

%% file: prooftree.tex
\newdimen\proofrulebreadth \proofrulebreadth=.05em
\newdimen\proofdotseparation \proofdotseparation=1.25ex
\newdimen\proofrulebaseline \proofrulebaseline=2ex
\newcount\proofdotnumber \proofdotnumber=3
\let\then\relax
\def\hfi{\hskip0pt plus.0001fil}
\mathchardef\squigto="3A3B
\newif\ifinsideprooftree\insideprooftreefalse
\newif\ifonleftofproofrule\onleftofproofrulefalse
\newif\ifproofdots\proofdotsfalse
\newif\ifdoubleproof\doubleprooffalse
\let\wereinproofbit\relax
\newdimen\shortenproofleft
\newdimen\shortenproofright
\newdimen\proofbelowshift
\newbox\proofabove
\newbox\proofbelow
\newbox\proofrulename
\def\shiftproofbelow{\let\next\relax\afterassignment\setshiftproofbelow\dimen0 }
\def\shiftproofbelowneg{\def\next{\multiply\dimen0 by-1 }%
\afterassignment\setshiftproofbelow\dimen0 }
\def\setshiftproofbelow{\next\proofbelowshift=\dimen0 }
\def\setproofrulebreadth{\proofrulebreadth}

\def\prooftree{
\ifnum  \lastpenalty=1
\then   \unpenalty
\else   \onleftofproofrulefalse
\fi
\ifonleftofproofrule
\else   \ifinsideprooftree
        \then   \hskip.5em plus1fil
        \fi
\fi
\bgroup
\setbox\proofbelow=\hbox{}\setbox\proofrulename=\hbox{}%
\let\justifies\proofover\let\leadsto\proofoverdots\let\Justifies\proofoverdbl
\let\using\proofusing\let\[\prooftree
\ifinsideprooftree\let\]\endprooftree\fi
\proofdotsfalse\doubleprooffalse
\let\thickness\setproofrulebreadth
\let\shiftright\shiftproofbelow \let\shift\shiftproofbelow
\let\shiftleft\shiftproofbelowneg
\let\ifwasinsideprooftree\ifinsideprooftree
\insideprooftreetrue
\setbox\proofabove=\hbox\bgroup$\displaystyle 
\let\wereinproofbit\prooftree
\shortenproofleft=0pt \shortenproofright=0pt \proofbelowshift=0pt
\onleftofproofruletrue\penalty1
}

\def\eproofbit{
\ifx    \wereinproofbit\prooftree
\then   \ifcase \lastpenalty
        \then   \shortenproofright=0pt  
        \or     \unpenalty\hfil         
        \or     \unpenalty\unskip       
        \else   \shortenproofright=0pt  
        \fi
\fi
\global\dimen0=\shortenproofleft
\global\dimen1=\shortenproofright
\global\dimen2=\proofrulebreadth
\global\dimen3=\proofbelowshift
\global\dimen4=\proofdotseparation
\global\count255=\proofdotnumber
$\egroup  
\shortenproofleft=\dimen0
\shortenproofright=\dimen1
\proofrulebreadth=\dimen2
\proofbelowshift=\dimen3
\proofdotseparation=\dimen4
\proofdotnumber=\count255
}

\def\proofover{
\eproofbit 
\setbox\proofbelow=\hbox\bgroup 
\let\wereinproofbit\proofover
$\displaystyle
}%
\def\proofoverdbl{
\eproofbit 
\doubleprooftrue
\setbox\proofbelow=\hbox\bgroup 
\let\wereinproofbit\proofoverdbl
$\displaystyle
}%
\def\proofoverdots{
\eproofbit 
\proofdotstrue
\setbox\proofbelow=\hbox\bgroup 
\let\wereinproofbit\proofoverdots
$\displaystyle
}%
\def\proofusing{
\eproofbit 
\setbox\proofrulename=\hbox\bgroup 
\let\wereinproofbit\proofusing
\kern0.3em$
}

\def\endprooftree{
\eproofbit 
  \dimen5 =0pt
\dimen0=\wd\proofabove \advance\dimen0-\shortenproofleft
\advance\dimen0-\shortenproofright
\dimen1=.5\dimen0 \advance\dimen1-.5\wd\proofbelow
\dimen4=\dimen1
\advance\dimen1\proofbelowshift \advance\dimen4-\proofbelowshift
\ifdim  \dimen1<0pt
\then   \advance\shortenproofleft\dimen1
        \advance\dimen0-\dimen1
        \dimen1=0pt
        \ifdim  \shortenproofleft<0pt
        \then   \setbox\proofabove=\hbox{%
                        \kern-\shortenproofleft\unhbox\proofabove}%
                \shortenproofleft=0pt
        \fi
\fi
\ifdim  \dimen4<0pt
\then   \advance\shortenproofright\dimen4
        \advance\dimen0-\dimen4
        \dimen4=0pt
\fi
\ifdim  \shortenproofright<\wd\proofrulename
\then   \shortenproofright=\wd\proofrulename
\fi
\dimen2=\shortenproofleft \advance\dimen2 by\dimen1
\dimen3=\shortenproofright\advance\dimen3 by\dimen4
\ifproofdots
\then
        \dimen6=\shortenproofleft \advance\dimen6 .5\dimen0
        \setbox1=\vbox to\proofdotseparation{\vss\hbox{$\cdot$}\vss}%
        \setbox0=\hbox{%
                \advance\dimen6-.5\wd1
                \kern\dimen6
                $\vcenter to\proofdotnumber\proofdotseparation
                        {\leaders\box1\vfill}$%
                \unhbox\proofrulename}%
\else   \dimen6=\fontdimen22\the\textfont2 
        \dimen7=\dimen6
        \advance\dimen6by.5\proofrulebreadth
        \advance\dimen7by-.5\proofrulebreadth
        \setbox0=\hbox{%
                \kern\shortenproofleft
                \ifdoubleproof
                \then   \hbox to\dimen0{%
                        $\mathsurround0pt\mathord=\mkern-6mu%
                        \cleaders\hbox{$\mkern-2mu=\mkern-2mu$}\hfill
                        \mkern-6mu\mathord=$}%
                \else   \vrule height\dimen6 depth-\dimen7 width\dimen0
                \fi
                \unhbox\proofrulename}%
        \ht0=\dimen6 \dp0=-\dimen7
\fi
\let\doll\relax
\ifwasinsideprooftree
\then   \let\VBOX\vbox
\else   \ifmmode\else$\let\doll=$\fi
        \let\VBOX\vcenter
\fi
\VBOX   {\baselineskip\proofrulebaseline \lineskip.2ex
        \expandafter\lineskiplimit\ifproofdots0ex\else-0.6ex\fi
        \hbox   spread\dimen5   {\hfi\unhbox\proofabove\hfi}%
        \hbox{\box0}%
        \hbox   {\kern\dimen2 \box\proofbelow}}\doll%
\global\dimen2=\dimen2
\global\dimen3=\dimen3
\egroup 
\ifonleftofproofrule
\then   \shortenproofleft=\dimen2
\fi
\shortenproofright=\dimen3
\onleftofproofrulefalse
\ifinsideprooftree
\then   \hskip.5em plus 1fil \penalty2
\fi
}
